\documentclass[a4paper]{amsart}

\usepackage{amsmath,amsfonts,amssymb}
\usepackage{graphicx}

\usepackage{hyperref}
\usepackage[T1]{fontenc}
\usepackage[utf8]{inputenc}
\usepackage{isabelle,isabellesym}

\newtheorem{theorem}{Theorem}
\theoremstyle{plain}

\newtheorem{corollary}{Corollary}

\newtheorem{definition}{Definition}

\newtheorem{lemma}{Lemma}

\numberwithin{equation}{section}

\let\ts=\thinspace

\begin{document}

\title[Szemer\'{e}di's Regularity Lemma and Roth's Theorem in Isabelle/HOL] 
    {Formalising Szemer\'{e}di's Regularity Lemma and Roth's Theorem on Arithmetic Progressions in Isabelle/HOL}

\author[Edmonds]{Chelsea Edmonds}
\author[Koutsoukou-Argyraki]{Angeliki Koutsoukou-Argyraki}
\author[Paulson]{Lawrence C. Paulson}
\email[Chelsea Edmonds]{cle47@cam.ac.uk}
\email[Angeliki Koutsoukou-Argyraki]{ak2110@cam.ac.uk}
\email[Lawrence C. Paulson]{lp15@cam.ac.uk}

\maketitle

\begin{center}
Department of Computer Science and Technology \\
University of Cambridge, UK
\end{center}

\begin{abstract}
We have formalised Szemer\'{e}di's Regularity Lemma and Roth's Theorem on Arithmetic Progressions, two major results in extremal graph theory and additive combinatorics, using the proof assistant Isabelle/HOL\@.
For the latter formalisation, we used the former to first show the Triangle Counting Lemma and the Triangle Removal Lemma: themselves important technical results. Here, in addition to showcasing the main formalised statements and definitions, we focus on sensitive points in the proofs, describing how we overcame the difficulties that we encountered.
\end{abstract}

\section{Introduction and Background} \label{sec:intro}

Szemer\'{e}di's Regularity Lemma and Roth's Theorem on Arithmetic Progressions are central results within extremal graph theory, additive combinatorics and, in a broader sense, number theory. They belong to a line of mathematical research which
finds its origins in Ramsey theory \cite{graham-ramsey-theory}: van der Waerden's Theorem, proved in 1927 and referring to arithmetic progressions, can be regarded as a direct precursor:
\begin{theorem}(van der Waerden) For any given $c,k \in \mathbb{N}$, there exists a number $N$ such that if the consecutive  integers $1$, $2$, $\ldots$,~$N$ are coloured, each with one of $c$ different colours, then there are at least $k$ integers in arithmetic progression whose elements are all of the same colour.
\end{theorem}
Less than a decade later, in 1936, Erd\H{o}s and Tur\'{a}n introduced a conjecture \cite{erdos-some-sequences} which was eventually proved in 1975 by Endre Szemer\'{e}di \cite{szemeredi-progression}---Gowers \cite{gowers-erdos-arithmetic} discusses the background to this result---and today is known as Szemer\'{e}di's Theorem:
\begin{theorem} \label{SzemerediTheorem}(Szemer\'{e}di)
 Every set of integers $A$ with positive upper asymptotic density contains a $k$-term arithmetic progression for every $k \in \mathbb{N}$.
\end{theorem}
The upper asymptotic density is a measure of the size of a set of integers. 
\begin{definition} The \emph{upper asymptotic density} of a set $A \subseteq \mathbb{Z}$ is defined as
$$\lim \sup_{N \rightarrow \infty}  \frac{|A \cap [1, N ]|}{N}.$$
\end{definition}
\noindent
For example, the set of even numbers has density $1/2$, while the set of primes has density zero. It can be shown that the set of square-free integers has density $6/\pi^2$, which tells us that ``most'' integers are square-free.

Szemer\'{e}di's original proof was combinatorial, but many further proofs were given,  most notably by Furstenberg in 1977 using ergodic theory \cite{furstenberg-ergodic-behavior} and by Gowers in 2001 using both Fourier analysis and combinatorics \cite{gowers-new-proof}.
It is worth mentioning that Szemer\'{e}di's Theorem is a fundamental ingredient in the proof of the (2004) celebrated Green--Tao Theorem  \cite{green-tao-primes}, which attests that the primes contain arbitrarily long arithmetic progressions. Although it does not directly follow from Szemer\'{e}di's Theorem, as the primes have zero asymptotic density in the integers, a Szemer\'{e}di-type statement plays a crucial rule in the proof, as noted by Conlon et al.~\cite{conlon-green-tao}.

\medskip
Already in 1953, Klaus Roth had shown a special case of Szemer\'{e}di's Theorem, proving the aforementioned 1936 conjecture by Erd\H{o}s and Tur\'{a}n \cite{erdos-some-sequences} for the case of arithmetic progressions of length $k=3$ \cite{roth-on-certain}.
His result, which is considered a milestone in additive combinatorics, is known as Roth's Theorem on Arithmetic Progressions:
\begin{theorem}(Roth) \label{RothTh} Every subset of the integers with positive upper asymptotic density contains a 3-term arithmetic progression.
\end{theorem}
Roth's original proof \cite{roth-on-certain} made use of Fourier analysis.
However, a later proof follows a combinatorial approach: it relies on Szemer\'{e}di's Regularity Lemma,
which Szemer\'{e}di proved in 1975 as a step towards his aforementioned Theorem \ref{SzemerediTheorem} \cite{diestel-graph,szemeredi-progression}.
Szemer\'{e}di first showed \cite{szemeredi-progression} a weaker version of the lemma for bipartite graphs, which was already sufficient to prove Theorem \ref{SzemerediTheorem}; later on, he showed the full lemma, for general graphs \cite{szemeredi-regular}.
Essentially, Szemer\'{e}di's Regularity Lemma attests that for any large dense graph, we can partition its vertices into a bounded number of parts, so that edges between most different parts behave in a ``random'' way.
To give a sense of what is meant by this notion of ``randomness'' on a quantitative level, we introduce the following definitions.

\medskip
In the following, if $G$ is a graph, $V(G)$ and $E(G)$ will denote the sets of its vertices and edges, respectively.
Sometimes the notation $G= (V, E)$ will be used.

For sets of vertices $X$, $Y \subseteq V(G)$, let $e (X, Y)$ be the number of edges between $X$ and $Y$.  That is, $$e (X,Y) = | \{ (x, y) \in X \times Y :\, xy \in E(G) \}|.   $$
\begin{definition} (Edge density) \label{def:edge-density}
Given a graph $G$, for sets of vertices $X, Y \subseteq V(G) $, we define the  edge density between
$X$ and $Y$ to be $$ d (X, Y) = \frac{e (X, Y)} {|X| |Y| }.$$
\end{definition}

\begin{definition} ($\epsilon$-regular pair) \label{def:regular-pair}
Given a graph $G$ and $\epsilon>0$, for sets of vertices $X$, $Y \subseteq V(G) $, we call $(X, Y)$ an $\epsilon$-regular pair (in $G$) if for all $A \subseteq X$, $ B \subseteq Y $ with $|A| \geq \epsilon |X| $, $|B| \geq \epsilon |Y| $, one has
$$|d(A, B) - d(X, Y)| \leq \epsilon.$$
\end{definition}
Taking the contrapositive: if the pair is not $\epsilon$-regular,
then the irregularity is witnessed by some $A \subseteq X$, $B \subseteq Y$ such that
$|A| \geq \epsilon |X|$,   $|B| \geq \epsilon |Y|$ and
$|d(A, B) - d(X, Y)|  > \epsilon$. 
We use \textit{not} $\epsilon$-\textit{regular} and $\epsilon$-\textit{irregular} interchangeably.

We are interested in partitions of a graph in which the number of irregular pairs is limited by the following formula:
\begin{definition} ($\epsilon$-regular partition) \label{def:regular-partition}
Given a graph $G$ and $\epsilon>0$, a partition $P = \{ V_1,\ldots, V_k \}$ of $V(G)$ is an $\epsilon$-regular
partition if
$$\sum_{\substack{(i, j) \in [k]^2\\ \text{\rm $(V_i, V_j)$ not $\epsilon$-regular}}} |V_i| |V_j| \leq \epsilon |V(G)|^2.  $$
  \end{definition}
We can now formally state Szemer\'{e}di's Regularity Lemma:
\begin{theorem}(Szemer\'{e}di)\label{SzemerediRegularity} For every $\epsilon >0$, there exists a constant $M$ such that
every graph has an $\epsilon$-regular partition of its vertex set into at most $M$ parts.
\end{theorem}

Szemer\'{e}di's Regularity Lemma has a number of significant applications that go far beyond the---already groundbreaking---proofs of Szemer\'{e}di's Theorem and Roth's Theorem: most notably, algorithmic applications within various areas of computer science.
It is considered the cornerstone of extremal graph theory. Szemer\'{e}di gives an accessible overview and some interesting historical details \cite{raussen-interview}. Gowers has obtained quantitative results on the lower bound on the cardinality of the induced $\epsilon$-regular partition \cite{gowers-lower}. Gowers \cite{gowers-quasirandomness-counting,gowers-hypergraph} and R\"{o}dl et al. \cite{frankl-extremal-problems,nagle-counting-lemma,rodl-regularity-lemma} have proved various extensions of Szemer\'{e}di's Regularity Lemma to hypergraphs. On a different note, Terence Tao has studied Szemer\'{e}di's Regularity Lemma from a probability theory and information theory perspective \cite{tao-szemeredis-revisited}.

This paper discusses our formalisations of Szemer\'{e}di's Regularity Lemma~\cite{Szemeredi_Regularity-AFP} and Roth's Theorem on Arithmetic Progressions~\cite{Roth_Arithmetic_Progressions-AFP} using Isabelle/HOL\footnote{\url{https://isabelle.in.tum.de}} \cite{isa-tutorial}.
Isabelle is a proof assistant (interactive theorem prover) supporting higher-order logic, among other formalisms. It is notable for its large library, the \textit{Archive of Formal Proofs} (AFP), containing hundreds of entries of formalised mathematics in addition to hundreds more on theoretical computer science and formal verification.
It offers powerful automation for both proving and disproving. Proofs can be written in a legible structured language called Isar.
As of the writing of this article, the AFP contains 22 entries classified under graph theory and 30 under combinatorics (some of these possibly overlapping). Within combinatorics, we can mention our work formalising design theory~\cite{edmonds-modular-first,Fishers_Inequality-AFP}
 and ordinal partition theory~\cite{dzamonja-formalising}. Notably, the aforementioned van der Waerden's Theorem was recently formalised
in Isabelle/HOL by Kreuzer and Eberl \cite{Van_der_Waerden-AFP}. 

For the formalisations described in this paper, we have built upon Noschinski's formalisation of the girth and chromatic number theorem~\cite{Girth_Chromatic-AFP}, as it defines the basics of graph theory starting with elementary concepts such as  \isa{ugraphs}, \isa{uedges}, \isa{uverts} for undirected graphs and the sets of edges and vertices thereof respectively. Vertices are seen as natural 
numbers and edges as sets of natural numbers, so they are of type \isa{nat} and \isa{nat set} respectively. This library was initially chosen as it additionally provided foundations in probabilistic reasoning on graphs, which may have been required had we only followed Zhao's proof \cite{zhao17}. While this was ultimately unnecessary, this simple formalisation of undirected graphs proved easier to work with in comparison to other more extensive graph libraries in Isabelle which focus on directed graphs~\cite{noschinskiGraphLibrary2015}, which in turn tend to complicate formal reasoning on undirected graphs.

Our contribution begins by formalising a proof of Szemer\'{e}di's Regularity Lemma, following Yufei Zhao's online notes for a course taught recently at MIT \cite{zhao17}---these are now being reworked into a book \cite{zhao-book-draft}---as well as online notes written by Paul Russell from a combinatorics course taught at Cambridge by Timothy Gowers in 2004~\cite{gowers-topics-combinatorics}.
This work is discussed in more detail in Section~\ref{sec:formalising-S}.
Building on our formalisation of Szemer\'{e}di's Regularity Lemma \cite{Szemeredi_Regularity-AFP} and following again the aforementioned set of notes supplemented by Bell et al.~\cite{bell-using-szemeredis}, we formalised the proofs of the
Triangle Counting Lemma and the Triangle Removal Lemma (Section~\ref{sec:triangle}). 
Finally, we used these to prove Roth's Theorem on Arithmetic Progressions (Section~\ref{sec:formalising-R}).
In Section~\ref{sec:difficulties}, we include a general discussion on our comments and observations, summarising what we learned through the formalisation process and focussing on the difficulties we encountered. Independently, and around the same time with us,
Ya\"{e}l Dillies and Bhavik Mehta (also at Cambridge but in the Mathematics Department) formalised the aforementioned results in the Lean theorem prover \cite{Dillies_Mehta_ITP_2022}. Their formalisations\footnote{\url{https://github.com/leanprover-community/mathlib/tree/szemeredi/src/combinatorics/szemeredi}} 
are pending full incorporation to mathlib, Lean's library of formalised mathematical proofs. We learned of their simultaneous work while we were halfway through our own formalisation.
We briefly compare the two approaches in Section~\ref{sec:Lean}.
Finally, Section~\ref{sec:concl} is a short conclusion.

We have written our proofs with care, trying to reveal the key insights, as we believe that formalised mathematics should not restrict to merely certifying claims, but should also clarify the proof ideas.
In this paper we present only highlights, hoping that any missing elements are self-explanatory.
Both of our formalisations can be found on the Archive of Formal Proofs (AFP) \cite{Roth_Arithmetic_Progressions-AFP,Szemeredi_Regularity-AFP}.
The formal material presented below has been edited to improve readability.

\section{Formalising Szemer\'{e}di's Regularity Lemma} \label{sec:formalising-S}

\subsection{Defining the energy}

We start by presenting our Isabelle formalisations of the notions defined in Section~\ref{sec:intro}. As mentioned, we build on the existing basic graph theory definitions defined by Noschinski \cite{Girth_Chromatic-AFP}. 

\textit{Edge density} is defined straightforwardly, following Def.\ts\ref{def:edge-density}  above.

\begin{isabelle}
\isacommand{definition}\isanewline
\ \ "edge\_density\ X\ Y\ G\ \isasymequiv \ card(all\_edges\_between\ X\ Y\ G)\ /\ (card\ X\ *\ card\ Y)"
\end{isabelle}

\smallskip

When speaking of $\epsilon$-regular pairs, note that $\epsilon$ is actually a parameter and that one might refer to, say, an $\epsilon/3$-regular pair.
Such a complicated syntax is achievable in Isabelle but at the cost of much syntactic hackery.
The formal version therefore accepts $\epsilon$ as an ordinary argument. This is our formalised version of  Def.\ts\ref{def:regular-pair}: 

\begin{isabelle}
\isacommand{definition}\ "regular\_pair\ X\ Y\ G\ \isasymepsilon\isanewline
\ \ \ \ \ \ \ \ \ \ \ \isasymequiv\ \isasymforall A\ B.\ A\ \isasymsubseteq \ X\ \isasymand \ B\ \isasymsubseteq \ Y\ \isasymand\isanewline
\ \ \ \ \ \ \ \ \ \ \ \ \ \ \ \ \ \ \ \ (card\ A\ \isasymge \ \isasymepsilon \ *\ card\ X)\ \isasymand \ (card\ B\ \isasymge \ \isasymepsilon \ *\ card\ Y)\ \isasymlongrightarrow \isanewline
\ \ \ \ \ \ \ \ \ \ \ \ \ \ \ \ \ \ \ \ \ \isasymbar edge\_density\ A\ B\ G\ -\ edge\_density\ X\ Y\ G\isasymbar \ \isasymle \ \isasymepsilon "
\end{isabelle}

\smallskip

The proofs will be concerned with partitions of the vertices of the given graph, \isa{G}. In particular, we will need to collect all $\epsilon$-irregular pairs among the members of any given partition, \isa{P}:

\begin{isabelle}
\isacommand{definition}\ "irregular\_set\ \isasymepsilon\ G P\isanewline
\ \ \ \ \ \ \ \ \ \ \ \isasymequiv\ \{(R,S)|R\ S.\ R\isasymin P\ \isasymand \ S\isasymin P\ \isasymand \ \isasymnot \ regular\_pair\ R\ S\ G\ \isasymepsilon \}"
\end{isabelle}
\smallskip

As mentioned above (Def.\ts\ref{def:regular-partition}), a \textit{regular partition} has ``relatively few'' irregular pairs, parameterised by \isa{\isasymepsilon}:

\begin{isabelle}
\isacommand{definition}\ "regular\_partition\ \isasymepsilon\ G P \isanewline
\ \ \ \ \ \ \ \ \ \ \ \isasymequiv\ partition\_on\ (uverts\ G)\ P\ \isasymand \isanewline
\ \ \ \ \ \ \ \ \ \ \ \ \ (\isasymSum (R,S)\ \isasymin \ irregular\_set\ \isasymepsilon \ G\ P.\ card\ R\ *\ card\ S)\isanewline
\ \ \ \ \ \ \ \ \ \ \ \ \ \ \isasymle \ \isasymepsilon \ *\ (card\ (uverts\ G))\isactrlsup 2"\end{isabelle}

\smallskip

We now formalise the key definitions referring to the \textit{energy} with respect to subsets and/or (a) partition(s) of a graph.
The notion of energy with respect to subsets of the vertices $U, W \subseteq V(G)$ is defined as follows: 

\begin{isabelle}
\isacommand{definition}\ "energy\_graph\_subsets\ U W G \isanewline
\ \ \ \ \ \ \ \ \ \ \ \isasymequiv\ card\ U *\ card\ W *\ (edge\_density U W G)\isactrlsup 2\ /\ (card\ (uverts\ G))\isactrlsup 2"
\end{isabelle}

Now, considering \textit{partitions} \isa{P}, \isa{Q} (instead of sets as above)
we define the following notion of energy. As we discuss at the end of Section~\ref{sec:difficulties}, instead of representing the partitions using indices for the parts, which was our first approach, in our final version of the formalisation we preferred to simply denote a partition as a set of sets, so the energy in terms of
partitions was eventually defined as follows:

\begin{isabelle}
\isacommand{definition}\ "energy\_graph\_partitions\ G\ P\ Q\isanewline
\ \ \ \ \ \ \ \ \ \ \ \isasymequiv\ \isasymSum R\isasymin P.\isasymSum S\isasymin Q.\ energy\_graph\_subsets\ R\ S\ G"
\end{isabelle}

Referring to a single partition of a single vertex set (which can be the entire vertex set of a graph) the \textit{energy of the partition} (also referred to as \textit{mean square density} ~\cite{gowers-topics-combinatorics}) is defined as follows:
 
\begin{isabelle}
\isacommand{abbreviation}\isanewline
\ \  "mean\_square\_density\ G\ P\ \isasymequiv \ energy\_graph\_partitions\ G\ P\ P"
\end{isabelle}

\subsection{Some elementary lemmas}  \label{subsec:lemmas}
Let us look at some of the consequences of the definitions introduced. As usual with interactive theorem proving, it is helpful to prove  a few trivial facts for every definition. Here are some of the more interesting results.

\smallskip

The following inequality concerns a partition \isa{P} with \isa{k} many parts of a vertex set \isa{V} of a finite graph and is proved by induction on \isa{k}. Although straightforward, the formal proof is over 30 lines long.
\begin{isabelle}
\isacommand{lemma}\ sum\_partition\_le:\isanewline
\ \ \isakeyword{assumes}\ "finite\_graph\_partition\ V\ P\ k"\ "finite\ V"\isanewline
\ \ \isakeyword{shows}\ "(\isasymSum R\isasymin P.\ \isasymSum S\isasymin P.\ real\ (card\ R\ *\ card\ S))\ \isasymle \ (real(card\ V))\isactrlsup 2"
\end{isabelle}

\noindent
This immediately yields the basic result that the mean square density is bounded by 1:

\begin{isabelle}
\isacommand{lemma}\ mean\_square\_density\_bounded:\ \isanewline
\ \ \isakeyword{assumes}\ "finite\_graph\_partition\ (uverts\ G)\ P\ k"\ "finite\ (uverts\ G)"\ \isanewline
\ \ \isakeyword{shows}\ "mean\_square\_density\ G\ P\ \isasymle \ 1"
\end{isabelle}

\smallskip

The following identity---relating the edge density of a graph \isa{G} with respect to its vertex sets \isa{U}, \isa{W} to the edge densities with respect to a partition \isa{P} of the vertex set \isa{U} into \isa{n} parts---ought to be straightforward, but the formal proof is nearly 50 lines, by induction on \isa{n}.

\begin{isabelle}
\isacommand{lemma}\ edge\_density\_partition:\isanewline
\ \ \isakeyword{assumes}\ "finite\_graph\_partition\ U\ P\ n"\isanewline
\ \ \isakeyword{shows}\ "edge\_density\ U\ W\ G\ =\ (\isasymSum X\isasymin P.\ edge\_density\ X\ W\ G\ *\ card\ X)/card\ U"
\end{isabelle}

\smallskip

This identity is used to prove a key lemma: that refining a partition of a vertex set cannot make the energy decrease.
We follow Gowers's combinatorial proof, which is based on a direct calculation~\cite{gowers-topics-combinatorics} and eschews probabilistic reasoning. 
(In contrast, Zhao's approach~\cite{zhao17, zhao-book-draft} reasons about expected value.)
The full version of the lemma considers partitions of two sets, but we save work by considering a partition of only one of the sets, then using symmetry to obtain the full result.

\begin{isabelle}
\isacommand{lemma}\ energy\_graph\_partition\_half:\isanewline
\ \ \isakeyword{assumes}\ "finite\_graph\_partition\ U P n"\isanewline
\ \ \isakeyword{shows}\ "card U *\ (edge\_density U W G)\isactrlsup 2\isanewline
\ \ \ \ \ \ \ \ \ \ \isasymle \ (\isasymSum R\isasymin P.\ card\ R\ *\ (edge\_density R W G)\isactrlsup 2)"
\end{isabelle}

Here, we combine the two halves allowing both sides to be partitioned.
The proof is straightforward (20 lines), using the previous result twice along with the commutativity of edge density.
The following lemma states that partitioning subsets of the vertex set cannot make the energy decrease.

\begin{isabelle}
\isacommand{proposition}\ energy\_graph\_partition\_increase:\isanewline
\ \ \isakeyword{assumes}\ "finite\_graph\_partition U P k"\ \isanewline
\ \ \ \ \ \ \isakeyword{and}\ "finite\_graph\_partition W Q l"\isanewline
\ \ \isakeyword{shows}\ "energy\_graph\_partitions G P Q \isasymge \ energy\_graph\_subsets U W G"
\end{isabelle}

In a similar spirit, the following result attests that refining partitions further cannot make the energy decrease (here
partition \isa{Q} refines partition \isa{P} of the vertex set \isa{V} while partition \isa{Q'} refines partition \isa{P'} of the vertex set \isa{V'}) :
\begin{isabelle}
\isacommand{proposition}\ energy\_graph\_partitions\_increase:\isanewline
\ \ \isakeyword{assumes}\ "refines\ V\ Q\ P"\ "refines\ V'\ Q'\ P'"\ \isanewline
\ \ \ \ \ \ \isakeyword{and}\ "finite\ V"\ "finite\ V'"\ \isanewline
\ \ \isakeyword{shows}\ "energy\_graph\_partitions\ G\ Q\ Q'\ \isasymge \ energy\_graph\_partitions\ G\ P\ P'"
\end{isabelle}

The following result is a special case of the above for a single partition:

\begin{isabelle}
\isacommand{corollary}\ mean\_square\_density\_increase:\isanewline
\ \ \isakeyword{assumes}\ "refines\ V\ Q\ P"\ "finite\ V"\isanewline
\ \ \isakeyword{shows}\ "mean\_square\_density\ G\ Q\ \isasymge \ mean\_square\_density\ G\ P"
\end{isabelle}

\subsection{The Energy Boost Lemma}  \label{subsec:boost}
Having explored how the energy behaves with respect to partitioning subsets and to the further refining of partitions, 
we are ready to state the key Energy Boost Lemma ~\cite{zhao17,zhao-book-draft}:
for a graph $G$, given a pair of vertex sets $(U, W)$  that is \textit{not} $\epsilon$-regular and
 where the irregularity is witnessed by the pair $(U', W')$ where
 $U'\subseteq U$ and $W'\subseteq W$, we partition $U$ as $\{U',\,U\setminus U'\}$ and $W$ as $\{W',\, W\setminus W'\}$ and the energy increases by at least $$ \frac{\epsilon^4~|U|~|W|}{|V(G)|^2.} $$

The possibility that $U'=U$ or $W'=W$---not treated in any of our sources, as they all assumed the strict subset relation---slightly complicates the statement of the lemma.
We must introduce the function \isa{P2} to deal with degenerate partitions, ensuring that the empty set is never a member of a partition.

\begin{isabelle}
\isacommand{definition}\ "P2\ X\ Y\ \isasymequiv \ if\ X\ \isasymsubset \ Y\ then\ \{X,Y-X\}\ else\ \{Y\}"
\end{isabelle}

The proof is a messy 80 lines. Most of the effort goes into manipulating complicated summations, which can be tricky to do formally.
Once again, Zhao~\cite{zhao17,zhao-book-draft} employs probabilistic arguments in order to compare energies.
We did not attempt that, preferring the simple calculation given by Gowers~\cite{gowers-topics-combinatorics}.

Note that the offending $\epsilon$-irregular pair $(U',W')$ is mentioned explicitly in the assumptions and conclusion.

\begin{isabelle}
\isacommand{proposition}\ energy\_boost:\isanewline
\ \ \isakeyword{fixes}\ \isasymepsilon ::real\ \isakeyword{and}\ U W G\isanewline
\ \ \isakeyword{defines}\ "alpha\ \isasymequiv \ edge\_density\ U W G"\isanewline
\ \ \isakeyword{defines}\ "u\ \isasymequiv \ \isasymlambda X\ Y.\ edge\_density\ X\ Y\ G\ -\ alpha"\isanewline
\ \ \isakeyword{assumes}\ "finite\ U"\ "finite\ W"\isanewline
\ \ \ \ \isakeyword{and}\ "U'\ \isasymsubseteq\ U"\ "W'\ \isasymsubseteq \ W"\ "\isasymepsilon \ >\ 0"\isanewline
\ \ \ \ \isakeyword{and}\ U':\ "card\ U'\ \isasymge \ \isasymepsilon \ *\ card U"\ \isakeyword{and}\ W':\ "card\ W'\ \isasymge \ \isasymepsilon \ *\ card\ W"\isanewline
\ \ \ \ \isakeyword{and}\ gt:\ "\isasymbar u\ U'\ W'\isasymbar \ >\ \isasymepsilon "\isanewline
\ \ \isakeyword{shows}\ "(\isasymSum A\ \isasymin \ P2\ U' U.\ \isasymSum B\ \isasymin \ P2\ W'\ W.\ energy\_graph\_subsets\ A\ B\ G)\isanewline
\ \ \ \ \ \ \ \ \ \ \isasymge \ energy\_graph\_subsets\ U W G\isanewline
\ \ \ \ \ \ \ \ \ \ \ \ +\ \isasymepsilon \isacharcircum 4\ *\ (card\ U *\ card\ W)\ /\ (card\ (uverts\ G))\isactrlsup 2"
\end{isabelle}

\subsection{Energy Boost Lemma for an irregular partition}  \label{subsec:refinement}

Having established the above result which refers to pairs that are not $\epsilon$-regular, we build on it to prove a statement referring to a \textit{partition} that is not $\epsilon$-regular, that is, a partition that has $\epsilon$-irregular pairs whose total size is too big (Def.\ts\ref{def:regular-partition}).
This crucial statement attests that for any $\epsilon$-irregular partition~\isa{P} of the vertices of \isa{G}, we can always find a refinement  \isa{Q} of \isa{P}
that increases the energy by at least $\epsilon^5$, a small but positive quantity.

\begin{isabelle}
\isacommand{proposition}\ exists\_refinement:\isanewline
\ \ \isakeyword{assumes}\ "finite\_graph\_partition\ (uverts\ G)\ P\ k"\ \isakeyword{and}\ "finite\ (uverts\ G)"\ \isanewline
\ \ \ \ \isakeyword{and}\ "\isasymnot \ regular\_partition\ \isasymepsilon \ G\ P"\ \isakeyword{and}\ "\isasymepsilon \ >\ 0"\isanewline
\ \ \isakeyword{obtains}\ Q\ \isakeyword{where}\ "refines\ (uverts\ G)\ Q\ P"\ \ \ \ \ \ \ \ \ \ \ \ \ \ \ \ \ \ \ \ \ \isanewline
\ \ \ \ \ \ \ \ \ \ \ \ \ \ \ \ \ \ \ "mean\_square\_density\ G\ Q\ \isasymge \ mean\_square\_density\ G\ P\ +\ \isasymepsilon \isacharcircum 5"\isanewline
\ \ \ \ \ \ \ \ \ \ \ \ \ \ \ \ \ \ \ "\isasymAnd R.\ R\isasymin P\ \isasymLongrightarrow \ card\ \{S\isasymin Q.\ S\ \isasymsubseteq \ R\}\ \isasymle \ 2\ \isacharcircum \ Suc\ k"\isanewline
\ \ \ \ \ \ \ \ \ \ \ \ \ \ \ \ \ \ \ "card\ Q\ \isasymle \ k\ *\ 2\ \isacharcircum \ Suc\ k"
\end{isabelle}

The formal proof is based on the Energy Boost Lemma and on lemmas on the energy behaviour
with respect to subsets, partitions and refinements thereof that were presented in Section~\ref{subsec:lemmas}. 
It spans about 300 lines:
\begin{itemize}
  \item About 50 lines for constructing the common refinement \isa{Q} of \isa{P}, using the previous Energy Boost Lemma and taking care to exploit symmetries.
  \item A further 30 lines for deriving some of its properties prior to proving the four claims in the theorem statement.
  \item Then 40 lines to show the first claim (that \isa{Q} refines partition \isa{P}).
  \item The second claim, about mean square density, requires more calculations involving summations and totals 90 lines.
  \item The third claim, a cap on the cardinality of the refinement of each member \isa{R} of partition \isa{P}, requires nearly 70 lines.
  \item The final claim, about the cardinality of \isa{Q}, is easy: under 15 lines.
\end{itemize}

\subsection{Proving Szemerédi's Regularity Lemma itself}

The task is now straightforward. Whenever we have a partition that is \textit{not} $\epsilon$-regular, we repeatedly apply the lemma above, each time obtaining a refinement of the previous partition and increasing the energy by at least $\epsilon^5$. The energy of any partition cannot exceed~1 (recall the lemma \isa{mean\_square\_density\_bounded} of Section~\ref{subsec:lemmas}), forcing termination after at most $\lceil\epsilon^{-5}\rceil$ iterations.

The formalisation of this argument is 75 lines long. Specifying the iterative construction---that at each step a new partition refines a previous one, that the energy increases and that the cardinality is bounded---seems to be unreasonably difficult.
The iteration is formalised as a function on natural numbers and the properties above are proved by induction.
It is tedious to reason about the existential claims made by the main lemma and that they continue to hold at the end.
There should be a more concise and elegant formal proof.

Crucially, the upper bound on the number of iterations is independent of the graph $G$.
It is given by a tower of exponentials, as is shown by iterating the previous lemma's bound on the size of the refined partition. We need the lemma $k\, 2^{k+1} \le 2^{2^k}$, and as its proof is a concise induction, we present it in full (Fig.\ts\ref{fig:le_tower_2}).

The main statement (Theorem \ref{SzemerediRegularity}) is formalised in Isabelle as follows:
\begin{isabelle}
\isacommand{theorem}\ Szemeredi\_Regularity\_Lemma:\isanewline
\ \ \isakeyword{assumes}\ "\isasymepsilon \ >\ 0"\isanewline
\ \ \isakeyword{obtains}\ M\ \isakeyword{where}\isanewline
\ \ \ "\isasymAnd G.\ card\ (uverts\ G)\ >\ 0\ \isasymLongrightarrow \ \isasymexists P.\ regular\_partition\ \isasymepsilon \ G\ P\ \isasymand \ card\ P\ \isasymle \ M"
\end{isabelle}

\begin{figure}[hbt]
\begin{isabelle}
\isacommand{lemma}\ le\_tower\_2:\ "k\ *\ (2\ \isacharcircum \ Suc\ k)\ \isasymle \ 2\isacharcircum (2\isacharcircum k)"\isanewline
\isacommand{proof}\ (induction\ k\ rule:\ less\_induct)\isanewline
\ \ \isacommand{case}\ (less\ k)\isanewline
\ \ \isacommand{show}\ ?case\ \isanewline
\ \ \isacommand{proof}\ (cases\ "k\ \isasymle \ Suc\ (Suc\ 0)")\isanewline
\ \ \ \ \isacommand{case}\ False\isanewline
\ \ \ \ \isacommand{define}\ j\ \isakeyword{where}\ "j\ =\ k\ -\ Suc\ 0"\isanewline
\ \ \ \ \isacommand{have}\ kj:\ "k\ =\ Suc\ j"\isanewline
\ \ \ \ \ \ \isacommand{using}\ False\ j\_def\ \isacommand{by}\ force\isanewline
\ \ \ \ \isacommand{then}\ \isacommand{have}\ \isasymsection :\ "(2\isacharcircum j\ +\ 3)\ \isasymle \ (2::nat)\ \isacharcircum \ k"\isanewline
\ \ \ \ \ \ \isacommand{by}\ (simp\ add:\ Suc\_leI le\_less\_trans not\_less\_eq\_eq numeral\_3\_eq\_3)\isanewline
\ \ \ \ \isacommand{have}\ "k\ *\ (2\ \isacharcircum \ Suc\ k)\ \isasymle \ 6\ *\ j\ *\ 2\isacharcircum j"\isanewline
\ \ \ \ \ \ \isacommand{using}\ False\ \isacommand{by}\ (simp\ add:\ kj)\isanewline
\ \ \ \ \isacommand{also}\ \isacommand{have}\ "\isasymdots \ \isasymle \ 6\ *\ 2\isacharcircum (2\isacharcircum j)"\isanewline
\ \ \ \ \ \ \isacommand{using}\ kj\ less.IH\ \isacommand{by}\ force\isanewline
\ \ \ \ \isacommand{also}\ \isacommand{have}\ "\isasymdots \ <\ 2\isacharcircum (2\isacharcircum j\ +\ 3)"\isanewline
\ \ \ \ \ \ \isacommand{by}\ (simp\ add:\ power\_add)\ \isanewline
\ \ \ \ \isacommand{also}\ \isacommand{have}\ "\isasymdots \ \isasymle \ 2\isacharcircum 2\isacharcircum k"\isanewline
\ \ \ \ \ \ \isacommand{by}\ (simp\ add:\ \isasymsection)\isanewline
\ \ \ \ \isacommand{finally}\ \isacommand{show}\ ?thesis\isanewline
\ \ \ \ \ \ \isacommand{by}\ simp\ \ \ \ \ \ \isanewline
\ \ \isacommand{qed}\ (auto\ simp:\ le\_Suc\_eq)\isanewline
\isacommand{qed}%
\end{isabelle}
  \caption{Statement and proof that $k\, 2^{k+1} \le 2^{2^k}$} \label{fig:le_tower_2}
\end{figure}

\section{The Triangle Counting Lemma and the Triangle Removal Lemma} \label{sec:triangle}

Triangles have long been valuable tools in graph theory, particularly in the context of extremal and probabilistic combinatorics. While for our purposes, the Triangle Counting Lemma and the Triangle Removal Lemma were required for the proof of Roth's Theorem, they also have numerous other applications. Hence, the formalisation of these lemmas is a valuable contribution in their own right. For both the Triangle Counting Lemma and Triangle Removal Lemma we use a mix of Zhao's notes \cite{zhao17} which clearly outlines the main intuition behind the proof, complemented by Bell and Grodzicki's notes \cite{bell-using-szemeredis} which provide additional detail on the exact calculations which take place.

\subsection{Triangle definitions}
We begin with some definitions. Firstly, we formalise the idea of a triangle in a graph:
\begin{isabelle}
\isacommand{definition}\ "triangle\_in\_graph\ x\ y\ z\ G\ \isanewline
\ \ \ \ \ \ \ \ \ \ \ \isasymequiv\ (\{x,y\}\ \isasymin \ uedges\ G)\ \isasymand \ (\{y,z\}\ \isasymin \ uedges\ G)\ \isasymand \ (\{x,z\}\ \isasymin \ uedges\ G)"
\end{isabelle}

A triangle-free graph is simply defined as one where there exist no such \isa{x}, \isa{y}, and \isa{z} satisfying the above definition. We also define the set of all triangles formed by taking vertices from three (not necessarily distinct) sets:
\begin{isabelle}
\isacommand{definition}\ "triangle\_triples\ X\ Y\ Z\ G\isanewline
\ \ \ \ \ \ \ \ \ \ \ \isasymequiv\ \{(x,y,z)\ \isasymin \ X\ \isasymtimes \ Y\ \isasymtimes \ Z.\ triangle\_in\_graph\ x\ y\ z\ G\}"
\end{isabelle}

Note that the triangle definition assumes that the \textit{well-formed} assumption holds between \isa{uedges} and \isa{uverts}: that every edge of \isa{G} joins two vertices of~\isa{G}\@. 
The \isa{triangle\_in\_graph} definition can also be formally reasoned on using the alternative \isa{neighbor\_in\_graph} definition to capture that assumption.

\begin{isabelle}
\isacommand{definition}\ "neighbor\_in\_graph\ x\ y\ G\isanewline
\ \ \ \ \ \ \ \ \ \ \ \isasymequiv\ (x\ \isasymin \ uverts\ G\ \isasymand \ y\ \isasymin \ uverts\ G\ \isasymand \ \{x,y\}\ \isasymin \ uedges\ G)"
\end{isabelle}

It can clearly be seen that for the definitions above, the ordering of the vertices of the vertex set will not affect the result of either definition. However, we do note that based on the \isa{triangle\_triples} definition, if the sets \isa{X}, \isa{Y} and \isa{Z} are not disjoint, a triangle may appear more than once (using a different ordering). This is in line with the proof of the Triangle Counting Lemma in Zhao's notes \cite{zhao17}, which requires ordered triples.

However, this causes issues in later proofs where we are interested in counting the purely distinct triangles. In this case we define a function \isa{mk\_triangle\_set} to convert a triple to a set of size 3, and further define the \isa{triangle\_set}, which mirrors the \isa{triangle\_triples} definition but for unordered triples.

\subsection{Triangle Counting Lemma}
Using these definitions, we are now ready to formalise the Triangle Counting Lemma, which provides a minimum bound on the number of triangles in a graph.

\begin{lemma}(Triangle Counting Lemma) Given a graph $G$, let $X, Y, Z \subseteq V(G)$ so that $(X, Y), (Y, Z), (Z, X)$  are all $\epsilon$-regular pairs for some $\epsilon >0$.  Assuming that $d(X, Y), d(X, Z), d(Z, Y) \geq 2 \epsilon$,  the number of triples $(x, y, z) \in X \times Y \times Z$ such that $x, y, z $ form a triangle in $G$ is at least
 $$(1-2 \epsilon) (d(X,Y) - \epsilon)(d(X,Z) - \epsilon)(d(Y,Z) - \epsilon)  |X| |Y| |Z|.  $$
 \end{lemma}

The proof, as presented by Zhao~\cite{zhao17}, has four main components.
\begin{enumerate}
  \item Given a regular pair $(X,Y)$, we have an upper bound of $\epsilon |X|$ on the number of vertices in $X$ which have fewer than $(d(X,Y) - \epsilon)|Y|$ neighbours, i.e. which have a negligible neighbourhood size in $Y$.
  \item Using (1) on the regular pairs $(X,Y)$ and $(X, Z)$ from the lemma assumptions, we establish a lower bound on a subset of $X$ where all elements which meet the minimum bound on neighbourhood size in $Y$ and $Z$.
  \item We establish a lower bound for the number of edges between the neighbourhoods of $X$ in $Y$ and $Z$.
  \item We combine (2) and (3) to establish a lower bound on the total number of triangles in the graph.
\end{enumerate}

We first show (1) in the lemma \isa{regular\_pair\_neighbor\_bound}.
\begin{isabelle}
\isacommand{lemma}\ regular\_pair\_neighbor\_bound:\ \isanewline
\ \ \isakeyword{fixes}\ \isasymepsilon ::real\isanewline
\ \ \isakeyword{assumes}\ "finite\ (uverts\ G)"\isanewline
\ \ \isakeyword{assumes}\ "X\ \isasymsubseteq \ uverts\ G"\ \isakeyword{and}\ "Y\ \isasymsubseteq \ uverts\ G"\ \isakeyword{and}\ "card\ X\ >\ 0"\isanewline
\ \ \ \ \isakeyword{and}\ "uwellformed\ G"\ \isakeyword{and}\ "\isasymepsilon >0"\isanewline
\ \ \ \ \isakeyword{and}\ "regular\_pair\ X\ Y\ G\ \isasymepsilon "\ \isakeyword{and}\ "edge\_density\ X\ Y\ G\ \isasymge \ 2*\isasymepsilon "\isanewline
\ \ \isakeyword{shows}\ "card\{x\ \isasymin \ X.\ card\ (neighbors\_ss\ x\ Y\ G)\isanewline
\ \ \ \ \ \ \ \ \ \ \ \ \ \ \ \ \ \ \ \ <\ (edge\_density\ X\ Y\ G\ -\ \isasymepsilon)\ *\ card\ Y\}\ \ <\ \ \isasymepsilon \ *\ card\ X"
\end{isabelle}

The proof required a case split to first reason on the trivial case (not considered by any of our sources) where there are no vertices in \isa{X} meeting the negligible neighbourhood size condition. The main case proceeded by contradiction as described in our sources. Bell and Grodzicki's notes \cite{bell-using-szemeredis} proved valuable in this case, providing much more detail on the calculations taking place, which formed the basis of the proof. It should be noted that it was this proof which first raised the issue of the strict versus non-strict subset use in the regular pair definition, which we discuss further in Section~\ref{sec:difficulties}.

This lemma could now be used to perform (2) within the formal proof of the Triangle Counting Lemma. For (3), we establish a technical auxiliary lemma:

\begin{isabelle}
\isacommand{lemma}\ all\_edges\_btwn\_neighbor\_sets\_lower\_bound:\ \isanewline
\ \ \isakeyword{fixes}\ \isasymepsilon ::real\ \isanewline
\ \ \isakeyword{assumes}\ "X\ \isasymsubseteq \ uverts\ G"\ "Y\ \isasymsubseteq \ uverts\ G"\ "Z\ \isasymsubseteq \ uverts\ G"\isanewline
\ \ \ \ \isakeyword{and}\ "\isasymepsilon >0"\ \ "finite\ (uverts\ G)"\ "uwellformed\ G"\isanewline
\ \ \ \ \isakeyword{and}\ "finite\ X"\ "finite\ Y"\ "finite\ Z"\ \isanewline
\ \ \ \ \isakeyword{and}\ "regular\_pair\ X\ Y\ G\ \isasymepsilon"\ "regular\_pair\ Y\ Z\ G\ \isasymepsilon "\ "regular\_pair\ X\ Z\ G\ \isasymepsilon "\isanewline
\ \ \ \ \isakeyword{and}\ "edge\_density\ X\ Y\ G\ \isasymge \ 2*\isasymepsilon "\ "edge\_density\ X\ Z\ G\ \isasymge \ 2*\isasymepsilon "\isanewline
\ \ \ \ \ \ \ \ \ "edge\_density\ Y\ Z\ G\ \isasymge \ 2*\isasymepsilon "\isanewline
\ \ \ \ \isakeyword{and}\ "card\ (neighbors\_ss\ x\ Y\ G)\ \isasymge \ (edge\_density\ X\ Y\ G\ -\ \isasymepsilon )\ *\ card\ Y"\isanewline
\ \ \ \ \isakeyword{and}\ "card\ (neighbors\_ss\ x\ Z\ G)\ \isasymge \ (edge\_density\ X\ Z\ G\ -\ \isasymepsilon )\ *\ card\ Z"\isanewline
\ \ \ \ \isakeyword{and}\ "x\ \isasymin \ X"\isanewline
\ \isakeyword{shows}\ "card(all\_edges\_between\ (neighbors\_ss\ x\ Y\ G)\ (neighbors\_ss\ x\ Z\ G)\ G)\ \isanewline
\ \ \ \ \ \ \isasymge \ (edge\_density\ Y\ Z\ G\ -\ \isasymepsilon )\isanewline
\ \ \ \ \ \ \ *\ card\ (neighbors\_ss\ x\ Y\ G)\ *\ card\ (neighbors\_ss\ x\ Z\ G)"
\end{isabelle}

This requires some set-up in the proof, but is relatively straightforward.

Finally, (4) is completed within the proof of the \isa{triangle\_counting\_lemma}, for which we give the Isabelle lemma statement below.

\begin{isabelle}
  \isacommand{theorem}\ triangle\_counting\_lemma:\isanewline
  \ \ \isakeyword{fixes}\ \isasymepsilon ::real\ \isanewline
\ \ \isakeyword{assumes}\ "X\ \isasymsubseteq \ uverts\ G"\ "Y\ \isasymsubseteq \ uverts\ G"\ "Z\ \isasymsubseteq \ uverts\ G"\isanewline
\ \ \ \ \isakeyword{and}\ "\isasymepsilon >0"\ \ "finite\ (uverts\ G)"\ "uwellformed\ G"\isanewline
  \ \ \ \ \isakeyword{and}\ "regular\_pair\ X\ Y\ G\ \isasymepsilon"\ "regular\_pair\ Y\ Z\ G\ \isasymepsilon "\ "regular\_pair\ X\ Z\ G\ \isasymepsilon "\isanewline
\ \ \ \ \isakeyword{and}\ "edge\_density\ X\ Y\ G\ \isasymge \ 2*\isasymepsilon "\ "edge\_density\ X\ Z\ G\ \isasymge \ 2*\isasymepsilon "\isanewline
\ \ \ \ \ \ \ \ \ "edge\_density\ Y\ Z\ G\ \isasymge \ 2*\isasymepsilon "\isanewline
  \ \ \isakeyword{shows}\ "card\ (triangle\_triples\ X\ Y\ Z\ G)\isanewline
  \ \ \ \ \ \ \isasymge \ (1\ -\ 2*\isasymepsilon)\ *\ ((edge\_density\ X\ Y\ G)\ -\ \isasymepsilon )\ *\ ((edge\_density\ X\ Z\ G)\ -\ \isasymepsilon )\isanewline
\ \ \ \ \ \ \ \ \ \ \ \ \ \ \ \ \ \ \ *\ ((edge\_density\ Y\ Z\ G)\ -\ \isasymepsilon )\ *\ card\ X\ *\ card\ Y\ *\ card\ Z"
  \end{isabelle}

While the proof required a number of additional steps to manage sum and inequality manipulations, it was relatively straightforward to complete. Once again, these manipulations closely followed Bell and Grodzicki~\cite{bell-using-szemeredis}. While the level of detail in these notes was  helpful, the formalisation picked up on a number of minor errors in stages (3) and (4) in particular. For example, there was an \textit{and} instead of \textit{or} in one of the set definitions, a \textit{plus} instead of a \textit{minus} in one of the lower bound results, and in one summation the summation was presented to be over pairs of sets, rather than the cardinality of the edges between these sets.

\subsection{Triangle Removal Lemma}
The Triangle Removal Lemma is the first direct application of our formalisation of Szemer\'{e}di's Regularity Lemma, which was presented in Section~\ref{sec:formalising-S}. It gives a maximum bound on the number of triangles which must be removed such that a graph can be considered \textit{triangle-free}:

\begin{lemma}(Triangle Removal Lemma) For all $\epsilon>0$, there exists $\delta >0$ such that any graph on $N$ vertices with less than or equal to $\delta N^3$ triangles can be made triangle-free by removing at most $\epsilon N^2$ edges.
\end{lemma}

This lemma is frequently expressed in the language of Landau symbols as follows:
any graph $G$ on $N$ vertices with $o(N^3)$ triangles can be made triangle-free by removing $o(N^2)$ edges.
We chose to prove it in the concrete form above, since it was not clear how to formalise a proof of the Landau version. 

Zhao \cite{zhao17} presents an intuitive recipe for applying Szemer\'{e}di's Regularity Lemma to prove the Triangle Removal Lemma, which we mirror in our formalisation:

\begin{enumerate}
  \item \textit{Partition:} We use Szemer\'{e}di's Regularity Lemma to obtain an $\epsilon$-regular partition of the vertices.
  \item \textit{Clean:} We remove edges that ``behave poorly'' within the $\epsilon$-regular structure imposed. Specifically, this includes edges between irregular pairs, pairs with low edge density, and pairs where one part is small.
  \item \textit{Count:} We use the Triangle Counting Lemma to establish a contradiction and show that the ``cleaned'' graph is triangle-free.
\end{enumerate}

We first define the concepts of a regular graph, a dense graph, and a decent graph. 
These three collectively express that a given graph (with a partition of its vertex set) has been cleaned as described in Step~(2).
These definitions are used within the proof to improve readability and simplify reasoning.

A \textit{regular graph} has been partitioned such that all pairs are regular.
\begin{isabelle}
\isacommand{definition}\ "regular\_graph\ P\ G\ \isasymepsilon\isanewline
\ \ \ \ \ \ \ \ \ \isasymequiv \ \isasymforall R\ S.\ R\isasymin P\ \isasymlongrightarrow \ S\isasymin P\ \isasymlongrightarrow \ regular\_pair\ R\ S\ G\ \isasymepsilon "
\end{isabelle}

A \textit{dense graph} satisfies a minimum density for its non-empty edge sets.
\begin{isabelle}
\isacommand{definition}\ "edge\_dense\ X\ Y\ G\ \isasymepsilon\isanewline
\ \ \ \ \ \ \ \ \ \ \ \isasymequiv\ all\_edges\_between\ X\ Y\ G\ =\ \{\}\ \isasymor \ edge\_density\ X\ Y\ G\ \isasymge \ \isasymepsilon "\isanewline
\isacommand{definition} "dense\_graph\ P\ G\ \isasymepsilon \ \isasymequiv \ \isasymforall R\ S.\ R\isasymin P\ \isasymlongrightarrow \ S\isasymin P\ \isasymlongrightarrow \ edge\_dense\ R\ S\ G\ \isasymepsilon "
\end{isabelle}

A \textit{decent graph} satisfies a minimum size for partition members that are connected by at least one edge.
\begin{isabelle}
\isacommand{definition}\ "decent\ X\ Y\ G\ \isasymeta \isanewline
\ \ \ \ \ \ \ \ \ \ \ \isasymequiv\ all\_edges\_between\ X\ Y\ G\ =\ \{\}\ \isasymor \ (card\ X \isasymge \ \isasymeta \ \isasymand \ card\ Y\ \isasymge \ \isasymeta )"\isanewline
\isacommand{definition}\ "decent\_graph\ P\ G\ \isasymeta \ \isasymequiv \ \isasymforall R\ S.\ R\isasymin P\ \isasymlongrightarrow \ S\isasymin P\ \isasymlongrightarrow \ decent\ R\ S\ G\ \isasymeta "
\end{isabelle}

Additionally, we introduce a lemma to convert between a cardinality bound on our two triangle representations (ordered and unordered). This is essential after applying the Triangle Counting Lemma in the proof of the Triangle Removal Lemma, mentioned in Zhao's proof as the way of managing any ``overcounting'' which may occur.

\begin{isabelle}
\isacommand{lemma}\ card\_convert\_triangle\_rep:\isanewline
\ \ \isakeyword{assumes}\ "X\ \isasymsubseteq \ uverts\ G"\ \isakeyword{and}\ "Y\ \isasymsubseteq \ uverts\ G"\ \isakeyword{and}\ "Z\ \isasymsubseteq \ uverts\ G"\isanewline
\ \ \isakeyword{and}\ \ \ \ \ "finite\ (uverts\ G)"\ "uwellformed\ G"\isanewline
\ \ \isakeyword{shows}\ "card\ (triangle\_set\ G)\ \isasymge\isanewline
\ \ \ \ \ \ \ \ \ \ 1/6\ *\ card\ \{(x,y,z)\ \isasymin \ X\isasymtimes Y\isasymtimes Z.\ triangle\_in\_graph\ x\ y\ z\ G\}"
\end{isabelle}

We now present the Isabelle version of the Triangle Removal Lemma:

\begin{isabelle}
\isacommand{theorem}\ triangle\_removal\_lemma:\isanewline
\ \ \isakeyword{fixes}\ \isasymepsilon \ ::\ real\isanewline
\ \ \isakeyword{assumes}\ "\isasymepsilon \ >\ 0"\isanewline
\ \ \isakeyword{shows}\ "\isasymexists \isasymdelta ::real\ >\ 0.\ \isasymforall G.\ card(uverts\ G)\ >\ 0\ \isasymlongrightarrow \ uwellformed\ G\ \isasymlongrightarrow \ \isanewline
\ \ \ \ \ \ \ \ \ \ card\ (triangle\_set\ G)\ \isasymle \ \isasymdelta \ *\ card(uverts\ G)\ \isacharcircum \ 3\ \isasymlongrightarrow \isanewline
\ \ \ \ \ \ \ \ \ \ (\isasymexists G'.\ triangle\_free\_graph\ G'\ \isasymand \ uverts\ G'\ =\ uverts\ G\ \isasymand\isanewline
\ \ \ \ \ \ \ \ \ \ \ \ \ \ \ \ \ uedges\ G'\ \isasymsubseteq \ uedges\ G\ \isasymand \ \isanewline
\ \ \ \ \ \ \ \ \ \ \ \ \ \ \ \ \ card\ (uedges\ G\ -\ uedges\ G')\ \isasymle \ \isasymepsilon \ *\ (card\ (uverts\ G))\isactrlsup 2)"
\end{isabelle}

The formal proof first discharges the trivial case where \isa{\isasymepsilon~\isasymge~1}, when all edges can be deleted. This case is not considered explicitly in any of our sources, although the main proof requires \isa{\isasymepsilon~<~1}.

For the main case, we follow Zhao's recipe. The application of Szemer\'{e}di's  Regularity Lemma is straightforward, enabling us to obtain an upper bound \isa{M0} on a regular partition for any arbitrary graphs \isa{G}. We further define \isa{D0}, as a strict upper bound on \isa{\isasymdelta}, which is important in deriving a contradiction at the end of the proof. Following this application, we derive a number of useful facts on the partition which are used later in the proof.

Step (2) is where the formal proof begins to get complicated. For each of the classes of edges that ``behave poorly'', we define a variable representing the set of those edges, and establish an upper bound on the cardinality of each of these sets. This counting proved quite fiddly in a formal environment, reinforcing observations made during our previous work formalising counting proofs on combinatorial structures \cite{edmonds-modular-first}. As such, the \textit{clean} stage of our formal proof was significantly longer than the more intuitive reasoning used by both Zhao~\cite{zhao17} and Bell--Grodzicki~\cite{bell-using-szemeredis}.

The formal proof can now obtain a new graph excluding these edges. The final stage of our proof matches Step (3), showing that this cleaned graph must be triangle-free. Again, this required some fiddly counting reasoning using the bounds established in Step (2). To help structure this reasoning, we show that the new graph obtained is regular, dense, and decent (as per our earlier Isabelle definitions), with Bell and Grodzicki's notes proving particularly useful here. Having met these conditions, the Triangle Counting Lemma can now be applied and through the use of the \isa{card\_convert\_triangle\_rep} lemma we come to a contradiction and finish the proof as required.

\section{Formalising Roth's Theorem on Arithmetic Progressions} \label{sec:formalising-R}

We tackled this development in three stages: the Diamond-Free Lemma, then a technical lemma containing the main construction, and finally the result itself (Theorem \ref{RothTh}).
In this section, we show a few highlights of the formal proof.

\subsection{The Diamond-Free Lemma}

The Triangle Removal Lemma implies a key corollary, which in the literature  is often referred to as a Ruzsa-Szemer\'{e}di bound or the \textit{Diamond-Free Lemma}.
First we formalise the property of being a graph every edge of which  belongs to precisely one triangle:

\begin{isabelle}
\ \ "unique\_triangles\ G\isanewline
\ \ \ \ \isasymequiv \ \isasymforall e\ \isasymin \ uedges\ G.\ \isasymexists !T.\ \isasymexists x\ y\ z.\isanewline
\ \ \ \ \ \ \ \ \ \ \ \ \ \ \ \ \ \ \ \ \ T\ =\ \{x,y,z\}\ \isasymand \ triangle\_in\_graph\ x\ y\ z\ G\ \isasymand \ e\ \isasymsubseteq \ T"
\end{isabelle}

\smallskip
Now we can state the corollary.

\begin{corollary}\label{Diamond_free}
For all $\epsilon >0$, there exists a $N>0$, so that any graph $G$ with more than $N$ vertices  and such that every edge of $G$ lies in a unique triangle, we have that $| E(G)| \leq \epsilon |V(G)|^2.$
\end{corollary}

\begin{isabelle}
\isacommand{corollary}\ Diamond\_free:\isanewline
\ \ \isakeyword{fixes}\ \isasymepsilon \ ::\ real\ \isanewline
\ \ \isakeyword{assumes}\ "0\ <\ \isasymepsilon "\isanewline
\ \ \isakeyword{shows}\ "\isasymexists N>0.\ \isasymforall G.\ card(uverts\ G)\ >\ N\ \isasymlongrightarrow \ uwellformed\ G\ \isasymlongrightarrow \isanewline
\ \ \ \ \ \ \ \ \ \ \ unique\_triangles\ G\ \isasymlongrightarrow\ card\ (uedges\ G)\ \isasymle \ \isasymepsilon \ *\ (card\ (uverts\ G))\isactrlsup 2"
\end{isabelle}
The above claim can be rephrased in the language of Landau symbols as follows: 
given a graph $G$ on $N$ vertices so that every edge of $G$ lies in a unique triangle, $G$ has $o(N^2)$ edges.

Zhao offers a six-line proof of Corollary \ref{Diamond_free}, but the formal version, which does not follow Zhao's notation with Landau symbols, is well over a hundred lines.
It proceeds as follows.
Let $\epsilon>0$ be given.
Use the Triangle Removal Lemma with $\epsilon/3$ to obtain some suitable $\delta>0$ and then pick some integer $N \ge \frac{1}{3\delta}$.
Let $G=(V,E)$ be given such that $\vert V\vert > N$.
Half of the formal development goes to showing that (by the assumption of unique triangles) $G$ has exactly three times as many edges as it has triangles.
Thus, the number of triangles is bounded above by $\vert V\vert^2 / 3$ and therefore by $\delta\vert V\vert^3$.
Removing at most $(\epsilon/3)\,\vert V\vert^2$ edges from $G$ yields a triangle-free version $G'$.\
A triangle of $G$ clearly cannot involve any edges of $G'$, so the number of triangles in $G$ is bounded by the number of edges that were removed from $G$, from which $\vert E\vert \le \epsilon \vert V\vert^2$ follows.

The Isabelle proof is largely straightforward except regarding the unique triangles property and converting between the triangle $\{x,y,z\}$ and
the corresponding triplet of edges 
for the counting argument.
This is a typical example of a trivial fact (``three times as many edges as triangles'') that is cumbersome to formalise.

Corollary \ref{Diamond_free} will be employed in the proof of Theorem \ref{RothTh}. Its
statement and Isabelle formalisation are presented below.

\subsection{Roth's Theorem: the main argument}

We begin by defining 3-term arithmetic progressions.
The definition is polymorphic, and the formal development uses both natural number and integer versions.
\begin{isabelle}
\isacommand{definition}\ progression3\ ::\ "'a::comm\_monoid\_add\ \isasymRightarrow \ 'a\ \isasymRightarrow \ 'a\ set"\isanewline
\ \ \isakeyword{where}\ "progression3\ k\ d\ \isasymequiv \ \{k,\ k+d,\ k+d+d\}"
\end{isabelle}

\smallskip

Roth's theorem is equivalent to the statement that any set free of 3-term arithmetic progressions must be ``small'' in a certain sense:

\begin{theorem}(Roth) For every $\epsilon >0$, there exists a
$M\in \mathbb{N}$ so that for all $N \geq M$, for any subset of the naturals $A$ with
$A  \subseteq \{0, \ldots, N-1\}$, if $A$ does not contain a 3-term arithmetic progression, then $|A| < \epsilon N$.
\end{theorem}

Thus for any set $A$ as above, the cardinality of $A$ is $o(N)$, that is, $A$ is ``small''.
However, as before, we work in terms of a given $\epsilon>0$ rather than using Landau notation.

The Isabelle/HOL formalisation comprises nearly 500 lines. The formalised statement follows.

\begin{isabelle}
\isacommand{lemma}\ RothArithmeticProgressions\_aux:\isanewline
\ \ \isakeyword{fixes}\ \isasymepsilon ::real\isanewline
\ \ \isakeyword{assumes}\ "\isasymepsilon \ >\ 0"\isanewline
\ \ \isakeyword{obtains}\ M\ \isakeyword{where}\ "\isasymforall N\ \isasymge \ M.\ \isasymforall A\ \isasymsubseteq \ \{..<N\}.\isanewline
\ \ \ \ \ \ \ \ \ \ \ \ \ (\isasymnexists k\ d.\ d>0\ \isasymand \ progression3\ k\ d\ \isasymsubseteq \ A)\ \isasymlongrightarrow \ card\ A\ <\ \isasymepsilon \ *\ real\ N"
\end{isabelle}

As mentioned earlier, Corollary \ref{Diamond_free} (the Diamond-Free Lemma) will be employed in the proof. We start by taking  $A  \subseteq \{0, \ldots, N-1\}$ assuming that $A$ contains no
3-term arithmetic progression. We embed $A$ into a cyclic group: $A \subseteq \mathbb{Z} / M \mathbb{Z}$, where $M=2N+1$. We then construct  a tripartite graph $G$ so that each of its three parts is a copy of $\mathbb{Z} / M \mathbb{Z}$.  We then show that each edge of $G$ lies in exactly one triangle, and therefore by Corollary \ref{Diamond_free}  we get a bound on the number of edges of $G$, and thus, by construction, on the cardinality of $A$ too.

The formalisation of the tripartite graph~$G$ is interesting. We need to make three disjoint copies of the natural numbers below~$M$.
Since the vertices of a graph are already natural numbers, we use a bijection between $\mathbb{N} \times \mathbb{N}$ and $\mathbb{N}$.
The library function \isa{prod\_encode} maps a pair of natural numbers to a natural number, and \isa{prod\_decode} is its inverse.

The first function creates a part (vertex set) of~$G$ from a given label (0, 1 or 2) and the numbers below $M$. The other two return the label (or the original number below~$M$, respectively) given a vertex of~$G$.
\begin{isabelle}
\ \ \ \ \isacommand{define}\ part\_of\ \isakeyword{where}\ "part\_of\ \isasymequiv \ \isasymlambda \isasymxi .\ (\isasymlambda i.\ prod\_encode\ (\isasymxi ,i))\ `\ \{..<M\}"\isanewline
\ \ \ \ \isacommand{define}\ label\_of\_part\ \isakeyword{where}\ "label\_of\_part\ \isasymequiv \ \isasymlambda p.\ fst\ (prod\_decode\ p)"\isanewline
\ \ \ \ \isacommand{define}\ from\_part\ \isakeyword{where}\ "from\_part\ \isasymequiv \ \isasymlambda p.\ snd\ (prod\_decode\ p)"
\end{isabelle}

We prove some obvious identities relating these functions, and then define the three parts $X$, $Y$, $Z$ of~$G$:
\begin{isabelle}
\ \ \ \ \isacommand{let}\ ?X\ =\ "part\_of\ 0"\isanewline
\ \ \ \ \isacommand{let}\ ?Y\ =\ "part\_of\ (Suc\ 0)"\isanewline
\ \ \ \ \isacommand{let}\ ?Z\ =\ "part\_of\ (Suc\ (Suc\ 0))"
\end{isabelle}
Defining the edges of $G$ isn't easy. Zhao says (referring to the set $A$ above)
\begin{quotation}
  Connect a vertex $x \in X$ to a vertex $y \in Y$ if $y-x \in A$. Similarly, connect $z \in Z$ with $y \in Y$ if $z-y \in A$. Finally, connect $x \in X$ with $z \in Z$ if $(z-x)/2 \in A$. Because we picked $M$ to be odd, 2 is invertible modulo $M$ and this last step makes sense.
\end{quotation}
To formalise these difference relations, it seems easier to work in the type of integers. The function \isa{int} is the obvious embedding from the natural numbers.
Note that division by 2 has been expressed in terms of multiplication by $N+1$.
\begin{isabelle}
\ \ \ \ \isacommand{define}\ "diff\ \isasymequiv \ \isasymlambda a\ b.\ (int\ a\ -\ int\ b)\ mod\ (int\ M)"\isanewline
\ \ \ \ \isacommand{define}\ "diff2\ \isasymequiv \ \isasymlambda a\ b.\ ((int\ a\ -\ int\ b)\ *\ int(Suc\ N))\ mod\ (int\ M)"
\end{isabelle}
We need a dozen lines simply to prove this trivial fact (and more facts are needed):
\begin{isabelle}
\ \ \ \ \isacommand{have}\ "diff\ y\ x\ =\ int\ a\ \isasymlongleftrightarrow \ y\ =\ (x\ +\ a)\ mod\ M"\ \isakeyword{if}\ "y\ <\ M"\ "a\isasymin A"
\end{isabelle}

An auxiliary function captures the requirement that an edge set needs to connect specific parts of the tripartite graph satisfying a given difference relation:
\begin{isabelle}
\ \ \ \ \isacommand{define}\ Edges\ \isakeyword{where}\ "Edges\ \isasymequiv\ \isasymlambda X\ Y\ df.\ \isanewline
\ \ \ \ \ \ \ \ \{\{x,y\}|\ x\ y.\ x\isasymin X\ \isasymand \ y\isasymin Y\ \isasymand \ df(from\_part\ y)(from\_part\ x)\ \isasymin \ int`A\}"
\end{isabelle}

Finally, Zhao's definition of $G$ is straightforward:
\begin{isabelle}
\ \ \ \ \isacommand{define}\ XY\ \isakeyword{where}\ "XY\ \isasymequiv \ Edges\ ?X\ ?Y\ diff"\isanewline
\ \ \ \ \isacommand{define}\ YZ\ \isakeyword{where}\ "YZ\ \isasymequiv \ Edges\ ?Y\ ?Z\ diff"\isanewline
\ \ \ \ \isacommand{define}\ XZ\ \isakeyword{where}\ "XZ\ \isasymequiv \ Edges\ ?X\ ?Z\ diff2"\isanewline
\ \ \ \ \isacommand{define}\ G\ \isakeyword{where}\ "G\ \isasymequiv \ (?X\ \isasymunion \ ?Y\ \isasymunion \ ?Z,\ XY\ \isasymunion \ YZ\ \isasymunion \ XZ)"
\end{isabelle}

Unfortunately, that this construction satisfies the obvious properties is tricky even to formalise, let alone to prove. Consider the following claim:
\begin{isabelle}
\ \ \ \ \isacommand{have}\ uniq:\ "\isasymexists i<M.\ \isasymexists d\isasymin A.\ \isasymexists x\ \isasymin \ \{p,q,r\}.\ \isasymexists y\ \isasymin \ \{p,q,r\}.\ \isasymexists z\ \isasymin \ \{p,q,r\}.\ \isanewline
\ \ \ \ \ \ \ \ \ \ \ \ \ \ \ \ \ \ \ x\ =\ prod\_encode(0,\ i)\isanewline
\ \ \ \ \ \ \ \ \ \ \ \ \ \ \ \ \ \isasymand \ y\ =\ prod\_encode(1,\ (i+d)\ mod\ M)\isanewline
\ \ \ \ \ \ \ \ \ \ \ \ \ \ \ \ \ \isasymand \ z\ =\ prod\_encode(2,\ (i+d+d)\ mod\ M)"\isanewline
\ \ \ \ \ \ \isakeyword{if}\ T:\ "triangle\_in\_graph\ p\ q\ r\ G"\ \isakeyword{for}\ p\ q\ r
\end{isabelle}
It is a characterisation of an arbitrary triangle $\{p,q,r\}$ in~$G$.
The claim is that $p$, $q$, $r$ can be permuted as $x\in X$, $y\in Y$, $z\in Z$ so that there is one vertex in each of the three parts of the graph (in order!), and that $x$, $y$, $z$ encode the arithmetic progression $i$, $i+d$, $i+2d$ for $i<M$ and $d\in A$.
Zhao devotes two sentences to this claim. The formal proof takes more than 50 lines.
It takes us to a key milestone:
\begin{isabelle}
\ \ \ \ \isacommand{have}\ "unique\_triangles\ G"
\end{isabelle}
The proof that each edge of $G$ lies in a unique triangle is four sentences in Zhao's presentation and more than 180 lines  in Isabelle/HOL, requiring a case analysis with three quite similar proofs depending on the edge: \isa{e\ \isasymin \ XY}, \isa{e\ \isasymin \ YZ} or \isa{e\ \isasymin \ XZ}.

Zhao's proof \cite{zhao17} concludes as follows (\textit{Corollary 3.18} is our Corollary~\ref{Diamond_free}):
\begin{quotation}
Then Corollary 3.18 implies that $G$ has $o(M^2)$ edges. But by construction $G$ has precisely $3M \vert A\vert$ edges. Since $M = 2N+1$, it follows that $\vert A\vert$ is $o(N)$ as claimed.
\end{quotation}
We have 100+ lines of Isabelle to go. First, a simple proof that $\vert E \vert \le \epsilon/12\, \vert V\vert^2$:
\begin{isabelle}
\ \ \ \ \isacommand{have}\ *:\ "card\ (uedges\ G)\ \isasymle \ \isasymepsilon /12\ *\ (card\ (uverts\ G))\isactrlsup 2"\isanewline
\ \ \ \ \ \ \isacommand{using}\ X\ \isacartoucheopen X\ <\ card\ (uverts\ G)\isacartoucheclose \ \isacartoucheopen unique\_triangles\ G\isacartoucheclose \ \isacartoucheopen uwellformed\ G\isacartoucheclose\isanewline
\ \ \ \ \ \ \isacommand{by}\ blast
\end{isabelle}
Next, a result that will let us show that the edge sets \isa{XY}, \isa{YZ}, \isa{XZ} all have cardinality $M \vert A\vert$.
The defining relation is abstracted as~\isa{df}.
The proof takes some effort!
\begin{isabelle}
\ \ \ \ \isacommand{have}\ card\_Edges:\ "card\ (Edges\ (part\_of\ \isasymxi )\ (part\_of\ \isasymzeta )\ df)\ =\ M\ *\ card\ A"\isanewline
\ \ \ \ \ \ \isakeyword{if}\ "\isasymxi \ \isasymnoteq \ \isasymzeta "\ \isakeyword{and}\ df\_cancel:\ "\isasymforall a\isasymin A.\ \isasymforall i<M.\ \isasymexists j<M.\ df\ j\ i\ =\ int\ a"\ \isanewline
\ \ \ \ \ \ \ \ \ \ \ \ \ \ \ \ \ \isakeyword{and}\ df\_inj:\ "\isasymforall a.\ inj\_on\ (\isasymlambda x.\ df\ x\ a)\ \{..<M\}"\ \ \isakeyword{for}\ \isasymxi \ \isasymzeta \ df
\end{isabelle}

Having got this far, the rest is plain sailing. The edge sets are trivially shown to be disjoint, from which we obtain
$\vert E\vert = 3M \vert A\vert$ and therefore $\vert A\vert \le \epsilon N$. 
\begin{isabelle}
\ \ \ \ \isacommand{have}\ "card\ (uedges\ G)\ =\ 3\ *\ M\ *\ card\ A"\isanewline
\ \ \ \ \ \ \isacommand{by}\ (simp\ add:\ G\_def\ card\_Un\_disjnt)\isanewline
\ \ \ \ \isacommand{then}\ \isacommand{have}\ "card\ A\ \isasymle \ \isasymepsilon \ *\ (real\ M\ /\ 4)"\isanewline
\ \ \ \ \ \ \isacommand{using}\ *\ \isacartoucheopen 0\ <\ M\isacartoucheclose \ \isacommand{by}\ (simp\ add:\ cardG\ card\_edges\ power2\_eq\_square)\isanewline
\ \ \ \ \isacommand{also}\ \isacommand{have}\ "\isasymdots \ <\ \isasymepsilon \ *\ N"\isanewline
\ \ \ \ \ \ \isacommand{using}\ \isacartoucheopen N>0\isacartoucheclose \ \isacommand{by}\ (simp\ add:\ M\_def\ assms)\isanewline
\ \ \ \ \isacommand{finally}\ \isacommand{show}\ "card\ A\ <\ \isasymepsilon \ *\ N"\ \isacommand{.}
\end{isabelle}

\subsection{Roth's Theorem: the final version}

The version of  Roth's Theorem presented as Theorem \ref{RothTh} in Section~\ref{sec:intro}, that is, formulated using the notion of upper asymptotic density,
essentially constitutes the contrapositive of the lemma proved above: if  $A$ is in a certain sense ``big enough'' then it must contain a 3-term arithmetic progression.

\begin{isabelle}
\isacommand{theorem}\ RothArithmeticProgressions:\isanewline
\ \ \isakeyword{assumes}\ "upper\_asymptotic\_density\ A\ >\ 0"\isanewline
\ \ \isakeyword{shows}\ "\isasymexists k\ d.\ d>0\ \isasymand \ progression3\ k\ d\ \isasymsubseteq \ A"
\end{isabelle}
The notion of upper asymptotic density is in the development \textit{Ergodic Theory} from the Archive of Formal Proofs \cite{Ergodic_Theory-AFP}.
Assuming the negation of the conclusion, it is easy to contradict the assumption.

\section{Some Difficulties} \label{sec:difficulties}

Much of the effort in this project had not to do with the formalisation itself but with ascertaining precisely what to formalise. Although this material is considered mathematics of central importance, sources are conflicting about the basic definitions.

The first problematic definition is \textit{edge density}, Def.\ts\ref{def:edge-density}:
$$ d (X, Y) = \frac{e (X, Y)} {|X| |Y| }.$$
In one draft of his notes, Zhao mentions that the given definition of $e(X,Y)$ does not even equal the actual number of edges between $X$ and $Y$ unless those sets are disjoint.
So the question is whether to \textbf{require} $X$ and $Y$ to be disjoint. Many authors do, although Zhao and Gowers do not.
To see whether this omission was intentional, we examined the literature and easily found numerous sources of all kinds (lecture notes, preprints, slides and journal articles) requiring the sets to be disjoint. One specific example is Malliaris and Shelah \cite{malliaris-regularity-lemmas}. As already mentioned in Section~\ref{sec:intro},
Szemer\'{e}di originally proved his Regularity Lemma for bipartite graphs and then generalised it for arbitrary graphs:
this may be the source of discrepancy with respect to disjointness.
The question matters because it affects subsequent definitions, theorem statements and proofs.
Ultimately we decided to omit the constraint provisionally and were never forced to reimpose it.
In the video\footnote{\url{https://www.youtube.com/watch?v=vcsxCFSLyP8&t=939s}} of his MIT lecture,
Zhao clarifies that we are in principle allowed to include pairs $(V_i, V_j)$ with $i=j $ in the regular partition definition, 
Def.\ts\ref{def:regular-partition} (see around 12:45 in the video). This is what prompted us to omit the disjointness constraint both in the edge density within the regular pairs definition and in the regular partition definition, considering the more general case where $ i=j$ is allowed everywhere.

The next problematic definition was that of an $\epsilon$-regular pair, Def.\ts\ref{def:regular-pair}.
We call $(X, Y)$ an $\epsilon$-regular pair if a certain condition holds for all $A \subseteq X$ and $B \subseteq Y$.
However, both Gowers and Zhao specified strict subsets, $A \subset X$ and $B \subset Y$.
In this case, it seemed that there could be no doubt, because the Energy Boost Lemma requires strict subsets: it creates partitions $\{A,\, X\setminus A\}$ and $\{B,\, Y\setminus B\}$, and a component of a partition cannot be empty.
This definition worked for the formalisation of Szemer\'{e}di's Regularity Lemma.
Unfortunately, when we moved to the proof of Roth's Theorem, the version of the definition with strict subsets did not make sense.
Proving the Triangle Counting Lemma, at the very start we ``obtain a pair of subsets witnessing the irregularity of $(X,Y)$''
and one of these so-called subsets is $Y$ itself.
With a little effort, we were able to show that the two definitions of regular pair, strict and non-strict, coincide provided both $X$ and $Y$ contained at least two elements. This extremely weak but necessary proviso unfortunately introduced a degenerate case in the Triangle Counting Lemma that we could not prove. Instead we changed the definition of $\epsilon$-regular pair to involve non-strict subsets and redid the proof of the Regularity Lemma. The necessary correction to the Energy Boost Lemma introduced annoying but minor complications throughout the proof (in particular, the introduction of the function \isa{P2} to deal with degenerate partitions, as mentioned in Section~\ref{subsec:boost}). 
Eventually we learned that in combinatorics, $\subset$ and $\subseteq$ might be used interchangeably even within the same context, with $\subsetneq$ reserved for the strict form.

Another issue in the formalisation was how to represent partitions. All informal expositions write a partition as a family of sets indexed by natural numbers: $\{V_1,\ldots,V_k\}$.
The notation with indices looks natural and familiar.
The indexing plays a prominent role in the proofs: sometimes we refer to $(V_i, V_j)$ where $i<j$, so the order is also significant.
But as we refine such a partition, further partitioning each of the $V_i$, the task of assigning correct indices to each set is irksome.
So we---having completed the formalisation---redid it to formalise a partition as nothing but a set of sets.
The reworking did not take long and resulted in a slightly shorter and definitely clearer proof.
On the rare occasions when explicit indices were necessary, choosing an arbitrary ordering of the partition was sufficient.

On a related note, another difficulty was formalising the partition refinement step, the lemma Zhao calls \textit{Energy Boost for an irregular partition} (Section~\ref{subsec:refinement}).
Here, a partition $\{V_1,\ldots,V_k\}$ of the vertex set is given and for all $\epsilon$-irregular pairs $(V_i,V_j)$, a further partition of both members is induced by the Energy Boost Lemma.
The new partition must be a common, simultaneous refinement of all of those partitions. What must be done is fairly obvious but only to someone reasonably familiar with the material. (The latest drafts of Zhao's book cover these subtleties superbly.) The actual formalisation of the common refinement of a set of partitions (a set of set of sets) is the collection of all possible nonempty intersections involving a member of each of the partitions.
The idea is obvious enough but the formalisation contains a few tricky elements.

Finally, our sources differed on the maximum possible size of the partition of each $V_i$ mentioned above.
In the notes for Gowers's course \cite{gowers-topics-combinatorics} $2^{2k}$ is given, while according to the early version of the notes by Zhao \cite{zhao17} it is ~$2^k$.
We eventually discovered the updated version of Zhao's notes  \cite{zhao-book-draft} with the correct (depending on details of definitions) figure of $2^{k+1}$ and a hint that one must exploit symmetry
to avoid double counting $(V_i, V_j)$ and $(V_j, V_i)$ in order to fit within that bound. We have followed Zhao, who states that pairs where $ i=j$ are also included; we say more about the treatment of the diagonal in Section~\ref{sec:Lean} below.
The inequality given is $k\, 2^{k+1} \le 2^{2^k}$, which in the final induction delivers the required stack of exponentials.
Because in the notes for Gowers's course \cite{gowers-topics-combinatorics} a higher upper bound is given, this inequality is stated as $k\, 2^{2k} \le 2^{2^k}$, which however, is not true for $k=2$ (and Isabelle reports this counterexample unprompted). 
All three different aforementioned bounds for this lemma lead, however, to the same tower of exponentials, which Gowers~\cite{gowers-lower} proved to be tight.

In all these difficulties we have no one to blame but ourselves, since there were willing experts whom we could have consulted. Gowers works in a nearby department, and when we finally made contact with Zhao (having completed both formalisations) he was enthusiastic to help us clarify the ambiguity in the regular pair definition. And there is a further lesson: mathematicians expect the right methods to be used but are quite willing to overlook trivial details, while computer scientists expect everything to fit together perfectly. There is a difference in outlook that must somehow be bridged if the formalisation of mathematics is to become mainstream. At the same time,
we see that formalising mathematics with a proof assistant like Isabelle can be helpful in clarifying minor details and edge cases. This is not only because
the user is forced to examine every technical point while articulating a proof to a computer, but also because working with a formal proof can reveal delicate issues: for example, counterexample-finding tools implemented within Isabelle's automation may remind the user about missing assumptions and edge cases, or the users themselves may experiment to see where the proof breaks after minor modifications in the code.

\section{Independent formalisation in Lean}  \label{sec:Lean}

\def\clap#1{\hbox to 0pt{\hss#1\hss}}  
As noted in Section~\ref{sec:intro}, similar material was formalised in Lean by Ya\"{e}l Dillies and Bhavik Mehta around the same time \cite{Dillies_Mehta_ITP_2022} and
their formalisations\footnote{\url{https://github.com/leanprover-community/mathlib/tree/szemeredi/src/combinatorics/szemeredi}}
are pending full incorporation to mathlib, Lean's library of formalised mathematical proofs.

A notable difference between the two formalisations is that Dillies and Mehta  treated
the \textit{equitable} version of Szemer\'{e}di's Regularity Lemma, which yields an equitable partition of the vertex set.
A partition of a set of size $n$ into $k$ parts is equitable if every part has size $\lfloor n/k\rfloor$ or $\lceil n/k\rceil$.
In particular, the equitable version of Szemer\'{e}di's Regularity Lemma states that
\begin{theorem} \label{equitable} For every $\epsilon >0$ and $m_0$, there exists a constant $M$ such that
every graph $G$ has an $\epsilon$-regular equitable partition of its vertex set into $k$ parts with $m_0 \leq k \leq M$.
\end{theorem}
The proof is similar to the proof of the non-equitable version, but at every stage when the partition is refined (by the Energy Boost Lemma), a further refinement step is done to keep the new partition equitable.

We earlier noted that our sources suggested three different upper bounds on the size of the partition obtained via the Energy Boost Lemma for an irregular partition.
One of the three is numerically wrong, but the other two are both correct, depending on details of the definitions.
To clarify, recall Definition \ref{def:regular-partition}:
As we explained in the previous section, we removed the disjointness constraint both in the edge density within the regular pairs definition and in the regular partition definition, meaning that we considered the more general case where $ i=j$ is allowed everywhere.
 Dillies and Mehta also allow for pairs $(X, X)$ in the edge density definition, however in the regular partition definition, unlike us, they explicitly exclude
the possibility $ i=j$ (omitting the diagonal, explicitly ignoring all $(V_i, V_i)$ pairs), that is, in their version of
 Definition  \ref{def:regular-partition}  they instead consider the condition
$$\sum_{\substack{i\neq j\\(i, j) \in [k]^2\\ \clap{\text{\scriptsize\rm $(V_i, V_j)$ not $\epsilon$-regular}}}} |V_i| |V_j| \leq \epsilon |V(G)|^2.  $$
By omitting the diagonal pairs where $i=j$, the upper bound attained  in the Lean development is $2^k$ rather than $2^{k+1}$ as in our case.

The diagonal pairs can safely be ignored in the development by Dillies and Mehta, since they formalise the equitable version of Szemer\'{e}di's Regularity Lemma, Theorem \ref{equitable}:
if there are enough parts in the partition, then the proportion of pairs that are diagonal can be made small.

We are grateful to Timothy Gowers, who in a private email clarified this discrepancy between the two approaches. 
He moreover stated that he finds the non-equitable version that we formalised more mathematically natural:
e.g.\ if the graph is quasirandom, partitioning it arbitrarily into enough parts to allow ignoring the diagonal contributions looks artificial when you can just take a single part.
Gowers added that he is not aware of any practical applications where equitability would be required.

Dillies and Mehta followed a different route than we did from Szemer\'{e}di's Regularity Lemma to Roth's Theorem: via the Corners Theorem.
A \textit{corner} in $\mathbb{Z}^2$ is a three-element set of the form
$$\{(x,y), (x+d,y), (x,y+d)\}$$ with $d > 0$.
The Corners Theorem states that every corner-free subset of $[N]^2$ has size $o(N^2)$.
It has a short proof using the Triangle Removal Lemma and leads fairly directly to Roth’s Theorem.
As already sketched above, we followed a route via the Diamond-Free Lemma, Corollary \ref{Diamond_free}  (also referred to in the literature as a Ruzsa-Szemer\'{e}di bound).

Finally, it is worth mentioning that although the Isabelle/HOL type system is much simpler than Lean's (the latter uses dependent types), we never had to exercise any ingenuity in regard to types.

\section{Conclusions}  \label{sec:concl}

Szemer\'{e}di's Regularity Lemma and Roth's Theorem on Arithmetic Progressions are regarded as major results and our announcement of their formalisation was greeted enthusiastically \cite{buzzard_icm}. And yet, the formalisation was almost straightforward, the main difficulties stemming from ambiguities in our sources compounded by our unwise refusal to consult available experts. The formalisations are relatively short: about 1000 lines for Szemer\'{e}di's Regularity Lemma and 1500 for Roth's Theorem. Zhao's exposition of the two theorems takes up about six pages for each. A rough calculation yields a de Bruijn factor (the ratio of the sizes of the formalised material over the original material) of about four for both developments.
	This sort of mathematics is clearly suitable for formalisation, and in view of the minor inaccuracies we discovered in standard presentations, there is some value in doing so.

\subsection*{Acknowledgements}
Many thanks to Timothy Gowers and Yufei Zhao for valuable advice, and to Ya\"{e}l Dillies and Bhavik Mehta for fruitful discussions.
The referees scrutinised the manuscript with care and made numerous helpful suggestions.

\subsection*{Statements and declarations}
The authors were supported by the ERC Advanced Grant ALEXANDRIA (Project 742178). Edmonds is jointly funded by the Cambridge Trust (Cambridge Australia Scholarship) and a Cambridge Department of Computer Science Premium Research Studentship.

The authors have no financial or proprietary interests in any material discussed in this article.

For the purpose of open access, the authors have applied a Creative Commons Attribution (CC BY) licence to any Author Accepted Manuscript version arising from this submission.

\bibliographystyle{abbrv}
\bibliography{combinatorics}

\end{document}